\documentclass[runningheads]{llncs}

\usepackage{eccv}

\usepackage{eccvabbrv}

\usepackage{graphicx}
\usepackage{booktabs}
\usepackage{multirow, multicol}
\usepackage{adjustbox}
\usepackage{boldline}
\usepackage{float}
\usepackage{amsmath}
\usepackage{tikz}
\usepackage{marvosym}

\newcommand{\tikzxmark}{%
\tikz[scale=0.23] {
    \draw[line width=0.7,line cap=round] (0,0) to [bend left=6] (1,1);
    \draw[line width=0.7,line cap=round] (0.2,0.95) to [bend right=3] (0.8,0.05);
}}
\newcommand{\tikzcmark}{%
\tikz[scale=0.23] {
    \draw[line width=0.7,line cap=round] (0.25,0) to [bend left=10] (1,1);
    \draw[line width=0.8,line cap=round] (0,0.35) to [bend right=1] (0.23,0);
}}

\usepackage[accsupp]{axessibility}  

\usepackage[pagebackref,breaklinks,colorlinks,citecolor=eccvblue]{hyperref}

\usepackage{orcidlink}
\usepackage[symbol,hang,flushmargin]{footmisc}

\begin{document}



\title{CardiacNet: Learning to Reconstruct Abnormalities for Cardiac Disease Assessment from Echocardiogram Videos}

\author{
Jiewen Yang\inst{1} \and
Yiqun Lin\inst{1} \and
Bin Pu\inst{1} \and
Jiarong Guo\inst{1} \and
\\
Xiaowei Xu\inst{3}\textsuperscript{\Letter} \and
Xiaomeng Li\inst{1,2}\textsuperscript{\Letter}
}
\authorrunning{J, Yang et al.}

\institute{The Hong Kong University of Science and Technology \\
\email{\{jyangcu, ylindw, jguoaz\}@connect.ust.hk, \{eebinpu, eexmli\textsuperscript{\Letter}\}@ust.hk} \and
HKUST Shenzhen-Hong Kong Collaborative Innovation Research Institute, Futian, Shenzhen, China \and 
Guangdong Provincial People’s Hospital, Guangzhou, China \\
\email{xiao.wei.xu@foxmail.com\textsuperscript{\Letter}}}

\maketitle

\vspace{-14pt}
\begin{abstract}
Echocardiogram video plays a crucial role in analysing cardiac function and diagnosing cardiac diseases. Current deep neural network methods primarily aim to enhance diagnosis accuracy by incorporating prior knowledge, such as segmenting cardiac structures or lesions annotated by human experts. However, diagnosing the inconsistent behaviours of the heart, which exist across both spatial and temporal dimensions, remains extremely challenging. For instance, the analysis of cardiac motion acquires both spatial and temporal information from the heartbeat cycle. To address this issue, we propose a novel reconstruction-based approach 
named \textbf{CardiacNet} to learn a better representation of local cardiac structures and motion abnormalities through echocardiogram videos. CardiacNet accompanied by the \textbf{C}onsistency \textbf{D}eformation \textbf{C}odebook (CDC) and the \textbf{C}onsistency \textbf{D}eformed-\textbf{D}iscriminator (CDD) to learn the commonalities across abnormal and normal samples by incorporating cardiac prior knowledge. In addition, we propose benchmark datasets named \textbf{CardiacNet-PAH} and \textbf{CardiacNet-ASD} for evaluating the effectiveness of cardiac disease assessment. In experiments, our CardiacNet can achieve state-of-the-art results in three different cardiac disease assessment tasks on public datasets CAMUS, EchoNet, and our datasets. The code and dataset are available at: \href{https://github.com/xmed-lab/CardiacNet}
{https://github.com/xmed-lab/CardiacNet}
\vspace{-10pt}
\end{abstract}    
\section{Introduction}
\label{sec:intro}
\vspace{-6pt}
Echocardiogram video, being the most widely used and easily accessible imaging modality in the field of cardiac medicine, has been proposed as a valuable tool for assessing various cardiac diseases, such as congenital heart disease~\cite{mcleod2018echocardiography,lai2015echocardiography} and atypical cardiac motion~\cite{popp1976echocardiographic,upton1976echocardiographic,sanjeevi2023automatic}.
Currently, there are several artificial intelligence methods~\cite{ouyang2020video,ryser2022anomaly,ghorbani2020deep,sun2023chamber,pu2021fetal,liu2023deep,zaman2021spatio,pu2022mobileunet,lu2022yolox, pu2024hfsccd,tseng2024real,puunsupervised,pu2024m3,yang2023graphecho,zheng2023gl} available for the assessment and evaluation of cardiac conditions in echocardiography. For instance, EchoNet~\cite{ouyang2020video}, a state-of-the-art cardiac assessment method, employs an R2+1D network to extract global spatiotemporal features from echocardiogram videos for predicting ejection fraction (EF). However, while these methods excel at capturing spatiotemporal information, they tend to neglect the local characteristics of cardiac structure, specifically the cyclical heartbeat motion. Furthermore, their performance is still limited, which restricts their adaptability to a broader range of cardiac diseases.

\begin{figure}[!t]
    \centering
    \includegraphics[width=0.999\linewidth]{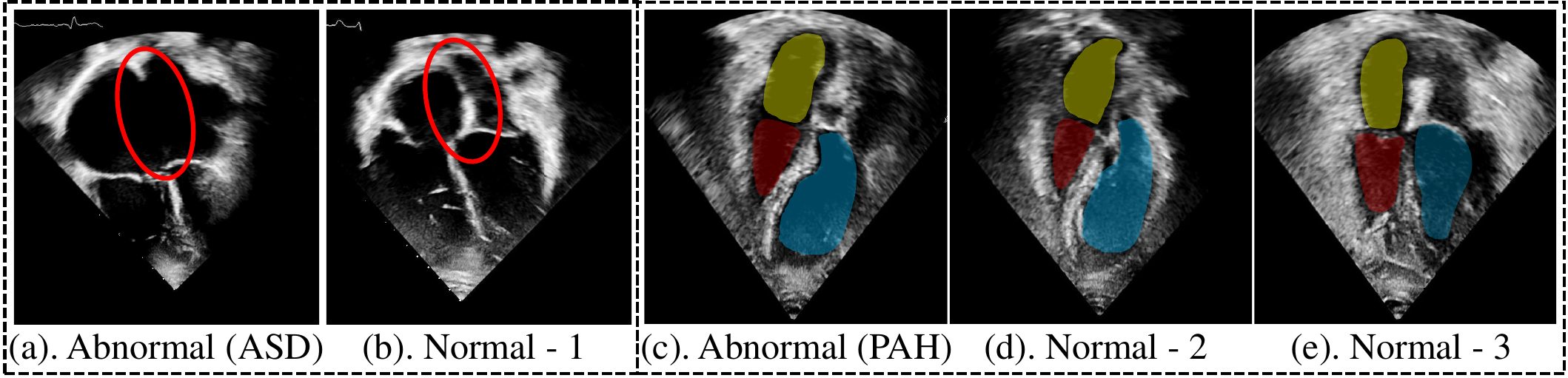}
    \vspace{-0.70cm}
    \caption{\textbf{Five examples from CardiacNet-PAH and CardiacNet-ASD datasets.} The appearance between the Atrial Septal Defect (ASD) (a) and normal (b) is easy to distinguish. For (c), (d) and (e), the appearance of cardiac structures in the Pulmonary Arterial Hypertension patient (c) and normal (d) are similar. In contrast, normal cases (d) and (e) show significant differences. Indicates that using a single image is not able to diagnose this type of cardiac disease. Clinically, experienced physicians will use echocardiogram videos with cardiac motion information to make diagnoses.}
    \label{fig:dataset_intrp}
    \vspace{-0.65cm}
\end{figure}

To develop a general approach for cardiac disease assessment, we have identified two important characteristics that encompass a wide range of common cardiac conditions, including EF, Pulmonary Arterial Hypertension (PAH), and Atrial Septal Defect (ASD).
Specifically, \textbf{(1) Local Structure Abnormality} refers to cardiac diseases that exhibit clear and distinctive abnormalities in a localized region within a single frame of an echocardiogram video. As depicted in Fig.~\ref{fig:dataset_intrp}(a-b), a hole (highlighted in {\color{red}{Red}}) can be observed in the atrial septum, enabling the mixing of blood between the left and right atria.
\textbf{(2) Cardiac Motion Abnormality} refers to cardiac diseases that may not have clear distinctive abnormalities in a single frame of echocardiogram videos, but can be detected through motion abnormalities of local cardiac structure observed in videos. For instance, in Fig.~\ref{fig:dataset_intrp}(c-d), there are no clear differences in cardiac structures between PAH patients and normal individuals based on a single frame of echocardiogram videos.
Therefore, it is highly necessary to develop an approach to learn a better representation across both temporal and spatial patterns of local cardiac structures via echocardiography.

%
%
%
%
%

Existing classification and regression-based disease assessment approaches~\cite{yu2022anatomy,ouyang2020video,mallya_2022_deepguide,huo2024hifuse} typically focus on global information and show difficulty in capturing local representations. 
In contrast, the reconstruction-based approaches~\cite{Ristea-CyTran-2023,ryser2022anomaly,silva2022constrained,kascenas2022denoising,VTGAN2021} offer a more intuitive solution by accurately reconstructing the abnormal and normal cases, enabling a deeper understanding of abnormality distribution, capturing fine-grained details, and achieving accurate disease assessment results.
However, existing reconstruction-based approaches were mainly designed for computed tomography (CT), magnetic resonance imaging (MRI), and X-ray modalities, focusing on abnormalities with low-level details such as tumors, bone fractures, and anomalous cardiac structures~\cite{ryser2022anomaly}. 
When directly applying these approaches to our datasets, their performance in assessing specific cardiac diseases with complex abnormalities is often unsatisfactory; see Tables~\ref{tab:results_EchoMD} and~\ref{tab:results_camus_echo}. 
This is mainly due to the fact that reconstructing abnormalities from echocardiogram videos is more challenging, as it requires considering the local structural and motion information presented by the heart.

To this end, we present a novel approach called CardiacNet for the assessment of various cardiac diseases. Our key assumption is that once the model is equipped with the capability to accurately reconstruct abnormalities from normal cases, it can gain a better understanding of the diseases in terms of their local structural details and motion changes, and vice versa. To achieve it, our CardiacNet consists of three key components: \textbf{(1)} \textbf{C}onsistency \textbf{D}eformation \textbf{C}odebook \textbf{(CDC)} is designed to simulate the reconstruction process between normal and abnormal cases, enabling the model to learn the local structural abnormalities and motion changes associated with the diseases. 
\textbf{(2)} \textbf{C}onsistency \textbf{D}eformation \textbf{D}iscriminator \textbf{(CDD)} aims to improve the quality of reconstructed videos and maintaining spatiotemporal consistency with the real videos in a discriminative manner. 
It prevents the degradation of reconstruction results by preserving the cardiac motion characteristics and introduces regional discrimination to maintain the local consistency of cardiac structural information.
\textbf{(3)} We introduce a bidirectional reconstruction network to facilitate the learning of feature distributions for both normal and abnormal cases. This approach enhances the reconstruction process, enabling us to establish the respective distributions and explicitly optimize the distributions of these two distinct groups.




%

%
%
\begin{table}[t!]
\setlength{\tabcolsep}{0.5pt}
\centering
  \caption{Summary statistics of datasets CardiacNet-PAH and CardiacNet-ASD and two public datasets CAMUS~\cite{leclerc2019deep} and EchoNet~\cite{ouyang2020video}.}
  \vspace{-11pt}
  \begin{adjustbox}{width=0.999\linewidth}
    \begin{tabular}{c|cccccc|cccccc}
\hlineB{3}
Dataset & \multicolumn{6}{c|}{CardiacNet-PAH (Ours)} & \multicolumn{6}{c}{CardiacNet-ASD (Ours)} \\ \hline
\multirow{2}{*}{\begin{tabular}[c]{@{}c@{}}Attri-\\ butes\end{tabular}} & \multicolumn{1}{c|}{\begin{tabular}[c]{@{}c@{}}Total \\ Videos\end{tabular}} & \multicolumn{1}{c|}{\begin{tabular}[c]{@{}c@{}}Total\\ Images\end{tabular}} & \multicolumn{1}{c|}{\begin{tabular}[c]{@{}c@{}}PAH \\ Cases\end{tabular}} & \multicolumn{1}{c|}{\begin{tabular}[c]{@{}c@{}}Normal\\ Cases\end{tabular}} & \multicolumn{1}{c|}{\begin{tabular}[c]{@{}c@{}}Other\\ Cases\end{tabular}} & \begin{tabular}[c]{@{}c@{}}Resol-\\ ution\end{tabular} & \multicolumn{1}{c|}{\begin{tabular}[c]{@{}c@{}}Total \\ Videos\end{tabular}} & \multicolumn{1}{c|}{\begin{tabular}[c]{@{}c@{}}Total\\ Images\end{tabular}} & \multicolumn{1}{c|}{\begin{tabular}[c]{@{}c@{}}ASD \\ Cases\end{tabular}} & \multicolumn{1}{c|}{\begin{tabular}[c]{@{}c@{}}Normal\\ Cases\end{tabular}} & \multicolumn{1}{c|}{\begin{tabular}[c]{@{}c@{}}Other\\ Cases\end{tabular}} & \begin{tabular}[c]{@{}c@{}}Resol-\\ ution\end{tabular} \\ \cline{2-13} 
 & \multicolumn{1}{c|}{496} & \multicolumn{1}{c|}{44,363} & \multicolumn{1}{c|}{342} & \multicolumn{1}{c|}{154} & \multicolumn{1}{c|}{0} & 720p & \multicolumn{1}{c|}{231} & \multicolumn{1}{c|}{13,471} & \multicolumn{1}{c|}{100} & \multicolumn{1}{c|}{131} & \multicolumn{1}{c|}{0} & 720p \\ \hline \hline \hlineB{2}
Dataset & \multicolumn{6}{c|}{CAMUS~\cite{leclerc2019deep}} & \multicolumn{6}{c}{EchoNet-Dynamic~\cite{ouyang2020video}} \\ \hline
\multirow{2}{*}{\begin{tabular}[c]{@{}c@{}}Attri-\\ butes\end{tabular}} & \multicolumn{1}{c|}{\begin{tabular}[c]{@{}c@{}}Total \\ Videos\end{tabular}} & \multicolumn{1}{c|}{\begin{tabular}[c]{@{}c@{}}Total\\ Images\end{tabular}} & \multicolumn{1}{c|}{\begin{tabular}[c]{@{}c@{}}EF$\geq$55\%\\ Cases\end{tabular}} & \multicolumn{1}{c|}{\begin{tabular}[c]{@{}c@{}}EF$\leq$50\%\\ Cases\end{tabular}} & \multicolumn{1}{c|}{\begin{tabular}[c]{@{}c@{}} 50\%<EF<55\%\\Cases\end{tabular}} & \begin{tabular}[c]{@{}c@{}}Resol-\\ ution\end{tabular} & \multicolumn{1}{c|}{\begin{tabular}[c]{@{}c@{}}Total \\ Videos\end{tabular}} & \multicolumn{1}{c|}{\begin{tabular}[c]{@{}c@{}}Total\\ Images\end{tabular}} & \multicolumn{1}{c|}{\begin{tabular}[c]{@{}c@{}}EF$\geq$55\%\\ Cases\end{tabular}} & \multicolumn{1}{c|}{\begin{tabular}[c]{@{}c@{}}EF$\leq$50\%\\ Cases\end{tabular}} & \multicolumn{1}{c|}{\begin{tabular}[c]{@{}c@{}}50\%<EF<55\%\\Cases\end{tabular}} & \begin{tabular}[c]{@{}c@{}}Resol-\\ ution\end{tabular} \\ \cline{2-13} 
 & \multicolumn{1}{c|}{500} & \multicolumn{1}{c|}{10,000} & \multicolumn{1}{c|}{201} & \multicolumn{1}{c|}{178} & \multicolumn{1}{c|}{121} & 480p & \multicolumn{1}{c|}{10,300} & \multicolumn{1}{c|}{1,755,250} & \multicolumn{1}{c|}{6961} & \multicolumn{1}{c|}{2246} & \multicolumn{1}{c|}{1093} & 120p \\ \hlineB{3}
\end{tabular}
  \end{adjustbox}
  \label{tab:detail_dataset}
  \vspace{-20pt}
\end{table}

We evaluate our method in EF prediction using two publicly available datasets, CAMUS~\cite{leclerc2019deep} and EchoNet~\cite{ouyang2020video}, which are the only publicly available echocardiogram video datasets for cardiac disease assessment. 
To comprehensively evaluate the performance of CardiacNet across a diverse array of cardiac diseases, we introduce two benchmark datasets, namely \textbf{CardiacNet-PAH} and \textbf{CardiacNet-ASD}, specifically designed for PAH and ASD assessment. 
A detailed comparison between our datasets and the public datasets is provided in Table~\ref{tab:detail_dataset}. Experimental results demonstrate that CardiacNet achieves state-of-the-art performance in three cardiac disease assessment tasks, including EF, PAH, and ASD.

To summarize, the main contributions of this paper are:
\begin{itemize}
    \item We have constructed two benchmark datasets, the CardiacNet-PAH and the CardiacNet-ASD, specifically designed for cardiac disease assessment using echocardiogram videos.
    \item CardiacNet is a novel approach that can capture local structural details and cardiac motion changes, enabling accurate assessment of cardiac diseases.
    \item Our CardiacNet surpasses prior work in classifying PAH and ASD with an improvement of 2.1\% and 5.0\% in accuracy. The CardiacNet also achieves a relative reduction of 5.2\% compared to prior arts in the EF prediction task.
\end{itemize}
\vspace{-12pt}
\section{Related Works}
\label{sec:related_works}
\subsection{Diseases Analysis on Different Modalities}
Currently, deep learning-based medical image representation learning on different modalities, such as CT, MRI and X-ray, typically use the reconstruction approach~\cite{Ristea-CyTran-2023,silva2022constrained,kascenas2022denoising,lin2024learning3dgaussiansextremely,lin2024c2rv,lin2023learning}. They usually learn the distribution from the control normal group and detect out-of-distribution abnormalities with significant low-level details, such as tumours and bone fractures. These approaches struggle to differentiate between the complex abnormalities of a specific disease, as the model focuses more on reconstructing each sample independently but lacks consideration across data samples. 
%
%
Gradient-weighted Class Activation Mapping~\cite{zhang2020viral,meena2023weakly,yu2022anatomy} can highlight the classification decision of feature maps from the network. Attention~\cite{sun2023chamber,schlemper2019attention} aims to highlight the out-of-distribution feature for abnormalities by introducing the attention mechanism. \cite{meena2023weakly,yu2022anatomy,zhang2020viral} use the anatomy-guided attention module to describe the confidence of the location of anomalies and treat them as explicit features to fine-tune the classification network. However, these methods rely on the classification backbone that is susceptible to noise and lacks the precision to accurately locate anomalous regions. 
The above methods serve for other medical modalities mainly focusing on medical images with significant lesions and pathology but lack consideration of both temporal and spatial information of cardiac data.
\vspace{-8pt}
\subsection{Cardiac Diseases Assessment from Echocardiogram Videos}
For echocardiogram video, the anomaly analysis can be grouped into anomalies classification~\cite{liu2023deep,lin2022echocardiography} and anomalies visualization~\cite{liu2023deep,ryser2022anomaly}, which offer baselines for adapting activation maps visualization of classification~\cite{liu2023deep}, and reconstruction-based~\cite{ryser2022anomaly} methods. \cite{lin2022echocardiography} first adapt the regional myocardial wall motion tracking to detect abnormalities and quantify cardiac function. However, it only focuses on a single cardiac structure and ignores other information. \cite{ryser2022anomaly} make the first attempt to reconstruct echocardiogram videos of normal groups from abnormal cases for congenital heart defect (CHD) detection. Yet this method barely considers prior knowledge of cardiac morphology. The lack of feature constraints also leads to the low quality of the reconstructed image. CAMUS~\cite{leclerc2019deep} and EchoNet-Dynamic~\cite{leclerc2019deep} are pioneer research that first proposes the echocardiogram video datasets for cardiac function evaluation. They also introduce the segmentation information for reference to predict the ejection fraction score. However, this task only reveals one of the cardiac functional parameters that is not able to classify the other cardiac diseases.

%
To overcome those problems, we thus propose a novel CardiacNet that builds a consistent relationship of morphological deformation between normal and abnormal cases by introducing prior knowledge of cardiac, which helps enable more accurate evaluation in different tasks. Our new CardiacNet-PAH and CardiacNet-ASD datasets provide two different cardiac diseases related to cardiac morphology abnormalities and motion dysfunction.
\vspace{-6pt}
\section{Methodology}
\begin{figure}[!t]
    \centering
    \includegraphics[width=0.999\linewidth]{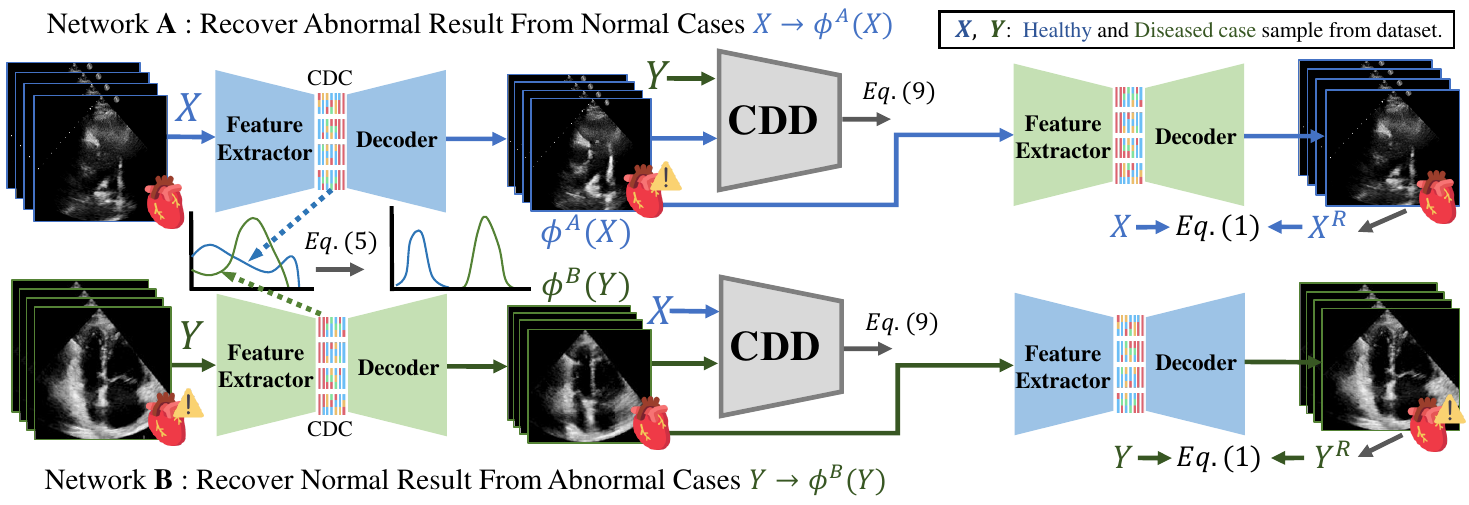}
    \vspace{-18pt}
    \caption{The overview of our CardiacNet, sample normal case $X$ and abnormal case $Y$, reconstruct the corresponding abnormal and normal results through networks $\phi^A(\cdot)$ and $\phi^B(\cdot)$, respectively. The Consistency Deformation Discriminator (CDD) is introduced to retain high reconstruction quality and allow the reconstruction results to be consistent with actual cases.}
    \label{fig:overview_pipeline}
    \vspace{-14pt}
\end{figure}

\vspace{-4pt}
    In this section, we introduce the CardiacNet as shown in Fig.~\ref{fig:overview_pipeline}. Hierarchically, CMT consists of the bidirectional reconstruction pipeline that simulates the deformation process from ``normal'' to ``abnormal'' cases and the reverse process. The \textbf{C}onsistency \textbf{D}eformation \textbf{C}odebook (CDC) is designed to formulate deformation processes, allows the network to identify patterns of cardiac structures and motion from data samples with a specific cardiac disease, expect reconstructed results to match the corresponding features of real samples. The introduction of module \textbf{C}onsistency \textbf{D}eformation \textbf{D}iscriminator (CDD) is to discriminate whether reconstructed results are consistent with real data samples both spatially and temporally. It also guarantees high-quality echocardiogram video reconstruction.
\vspace{-4pt}
\subsection{Bidirectional Reconstruction Network}
\label{sec:bidirectional_network}
As shown in Fig.~\ref{fig:overview_pipeline}, two independent networks $\phi^A(\cdot)$ and $\phi^B(\cdot)$ with the same type of feature extractor, deformation codebook and decoder, respond to the reconstruction process of cases between ``normal'' and ``abnormal''. Using the echocardiogram video input $X\in\mathbb{R}^{N\times H\times W\times 3}$ sample from the normal set as an example, where $N$ is the total frame number of the $X$. First, divide each frame of $X$ into regular non-overlapping patches and perform masking with the randomly sampled subset of patches. Then, compute reconstructed abnormal result $\phi^A(X)$. In the final, network $\phi^B(\cdot)$ transforms $\phi^A(X)$ as $X^R$ to the normal result as the same as input $X$. With the $L1$ loss as our supervised reconstruction loss as follows:
\begin{equation}
    \mathcal{L}_\text{recon}(X, X^R) = ||X-X^R||_1,
\end{equation}
where $||\cdot||_1$ indicate $L_1$ norm. The reconstructed abnormal $\phi^A(X)$ and real case sampled from the abnormal set will be discriminated by the CDD and compute the adversarial loss $\mathcal{L}_\text{adv}$. The process for reconstructing normal results from abnormal cases shares the same processing pipeline as the above description.
\vspace{-8pt}
\subsection{Consistency Deformation Codebook}
Section~\ref{sec:intro} and research~\cite{ganame2007regional,oxborough2012right,niemann2012echocardiographic,schafer2023myocardial,geske2014deformation} illustrated that human cardiac remains structurally and morphologically similar, the lesions of cardiac diseases and its motion are generally dominated by specific locations of main structures and their substructures (refer to Fig.~\ref{fig:dataset_intrp}). With a large number of medical cases confirmed by experts, the pattern of cardiac structures and motion between normal and abnormal are able to be learned across samples.
Hence, the main goal of the \textbf{C}onsistency \textbf{D}eformation \textbf{C}odebook (CDC) is designed to formulate the pattern from medical cases. We hypothesise that the network understands the representation of a specific disease that can also reconstruct normal from abnormal or its reverse process. Hence, to simulate such behaviours, 1). The proposed CDC constructs the regional representation for different cardiac structures in order to maintain the temporal and spatial properties consistent between original and reconstructed echocardiogram videos. 2). To differentiate the deformation from ``normal'' to ``abnormal'' and its reversed process, we use the transport distance to distribute the discrepancy of two different distributions from large data samples and optimize the CDC module of network $\phi^A(\cdot)$ and $\phi^B(\cdot)$.
%
%
\begin{figure*}[!t]
    \centering
    \includegraphics[width=0.999\linewidth]{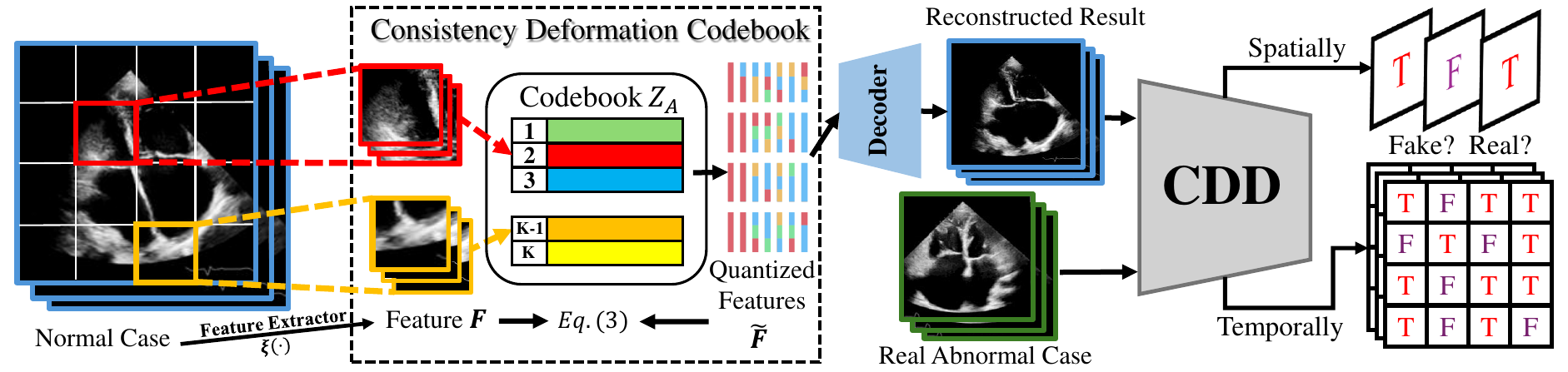}
    \vspace{-16pt}
    \caption{The description of the one-way process of our CardiacNet. The encoded feature $F$ of the normal case will be quantized by the deformation codebook $Z$ as quantized feature $\tilde{F}$. The decoder then recovers $\tilde{F}$ to the reconstructed abnormal result. Accompanied by the abnormal case sampled from the dataset, the Consistency Deformation Discriminator (CDD) is introduced to improve the consistency between reconstructed results and actual samples with regional discrimination.}
    \label{fig:overview_detail}
    \vspace{-14pt}
\end{figure*}
\\

\noindent
\textbf{Consistent Deformation Encoding.}
\label{sec:morphological_encoding}
As the pipeline described in Section~\ref{sec:bidirectional_network} and Fig.~\ref{fig:overview_detail}, the CDC receives the feature map $F$ encoded by the feature extractor $\xi(\cdot)$ of the network from the input.
As discussed above, formulating the deformation process regionally is a more natural fit in echocardiogram videos. In order to perform this approach, we discretise the continuous feature $F$ and reconstruct its latent representation regionally in a vector quantization manner. We first rewrite the $F$ as $F=\{F_{n,i,j}\}_{n,i,j}^{N\times h\times w}\subset\mathbb{R}^{d}$ for querying the codebook entries $\mathcal{Z}=\{Z_k\}_{k=1}^K\subset\mathbb{R}^{d}$, where $K$ is the total length of entries. In this step, directly applying the codebook~\cite{esser2021taming} for quantizing videos disrupts temporal consistency. Thus, we add learnable position encoding $\mathcal{P}=\{P_n\}_{n=1}^N\subset\mathbb{R}^{d}$ to feature maps $F$ along the temporal dimension, which guarantees temporal consistency locally and globally. 
Given a subsequent element-wise quantization $\sigma(\cdot)$, we generate reconstructed abnormal feature ${\sigma}({F})$ as following:
\begin{equation}
  \tilde{F}={\sigma}(F,\mathcal{Z},\mathcal{P}):=\left(\underset{Z_k \in \mathcal{Z}}{\arg \min}\Big\|(F_{n,i,j}+P_n)-Z_k\Big\|_2^2\right)_{n,i,j}\in\mathbb{R}^{t\times h\times w\times d}.
\label{eq:quantization}
\end{equation}

For the loss of CDC, following the previous research~\cite{esser2021taming,van2017neural}, we end-to-end train the CDC via Equation~\ref{eq:quantilize}.
\begin{equation}
  \mathcal{L}_{q}(\xi(I),\tilde{F}) = \Big\|sg[\xi(I)]-\tilde{F}\Big\|_2^2 + \lambda\cdot\Big\|sg\big[\tilde{F}\big]-\xi(I)\Big\|_2^2,
  \label{eq:quantilize}
\end{equation}
where $I, sg[\cdot]$, and $\lambda$ denote the input of network $\phi(\cdot)$, the stop-gradient operation, and the factor of the second loss item that is set as 0.25. Equation~\ref{eq:quantilize} guarantees the network commits to the $\mathcal{Z}$ since its dimensionless embedding space may grow arbitrarily during training. For the optimization of the CDC, we use the exponential moving average (EMA) method to update the codebook $\mathcal{Z}$ as the following equation:
\begin{equation}
    \mathcal{Z}'_{\text{new}} = (1-\omega)\cdot \mathcal{Z} + \omega\cdot \mathcal{Z}_{\text{new}},
    \label{eq:update_codebook}
\end{equation}
where $\omega$ is the weight for updating the current codebook that is set as 0.01.
\\

\noindent
\textbf{Optimal Transport Distance Optimization.}
The codebook of module CDC in Section~\ref{sec:morphological_encoding} is proposed to formulate the pattern of the deformation process through all data samples from the dataset. To distinguish the distribution of normal and abnormal sets that are learned by codebooks of network $\phi^A(\cdot)$ and $\phi^B(\cdot)$, 
a more intuitive way is to use relative entropy to represent how one probability distribution differs from another. In this paper, we adopt the optimal transport measurement and expect to maximize the distance of deformations between the normal and abnormal sets. 
\begin{figure}[!t]
    \centering
    \includegraphics[width=0.999\linewidth]{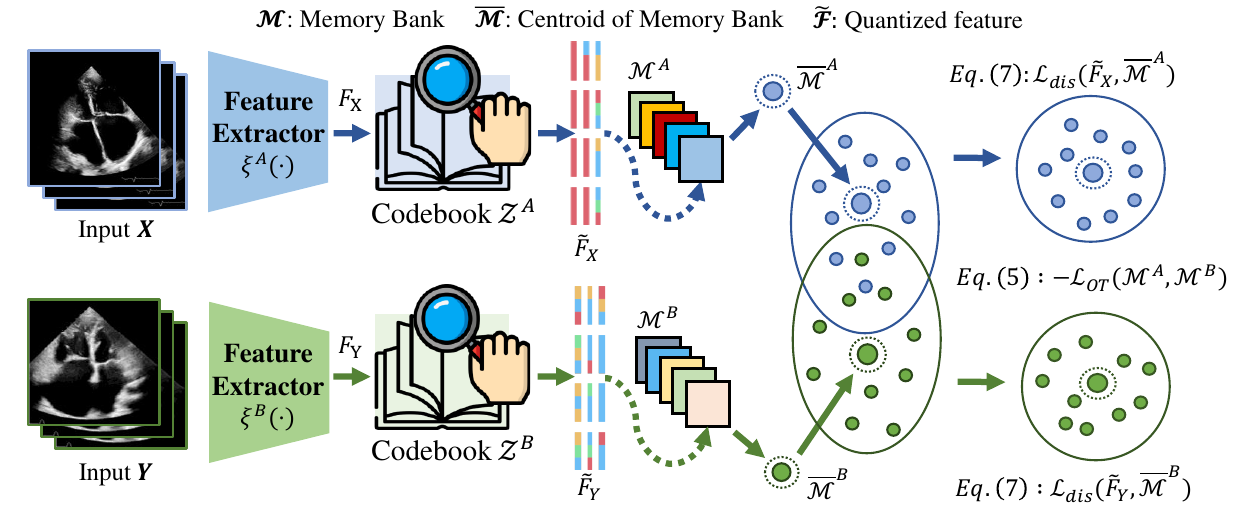}
    \vspace{-10pt}
    \caption{The optimal transport distance optimization between two networks $\phi^A(\cdot)$ and $\phi^B(\cdot)$. Memory banks $\mathcal{M}^A$ and $\mathcal{M}^B$ store the features of normal and abnormal data samples, respectively. The loss $\mathcal{L}_\text{OT}(\mathcal{M}^A,\mathcal{M}^B)$ makes these two distributions keep away from each other. Losses $\mathcal{L}_\text{dis}(\tilde{F}_{X},\overline{\mathcal{M}}^A)$, and $\mathcal{L}_\text{dis}(\tilde{F}_Y,\overline{\mathcal{M}}^B)$ make representations of clusters more consistent.}
    \label{fig:transport_optimal}
    \vspace{-10pt}
\end{figure}
As shown in Fig.~\ref{fig:transport_optimal}, networks $\phi^A(\cdot)$ and $\phi^B(\cdot)$ response for feature encoder, compute $F_X$, $F_Y$ from normal case $X$ and abnormal case $Y$. Implicitly, we can directly optimize the distance between codebooks of $\phi^A(\cdot)$ and $\phi^B(\cdot)$, which optimizes empirical distributions from entries instead of all data samples corresponding to each category from the dataset. 
However, due to the entries of each codebook being irrelevant and redundant, an entry in the same position of different codebooks is non-matching and non-equivalent, which can always be easily optimized to maximize their distance within a few iterations. Such orderless matching thus will invalidate the optimization, which indicates implicit optimization is not suitable for our approach.

To tackle this problem, we build two updated memory banks to store features encoded by CDC for normal and abnormal cases iteratively, which approximates the distribution of space of data samples. Similar to using EMA to update the codebook $\mathcal{Z}^A$ and $\mathcal{Z}^B$, in memory banks $\mathcal{M}^A$, $\mathcal{M}^B$, we replace the ancestral features with current descendant features from Equation~\ref{eq:quantization} and update the centroid with the EMA approach. Then, explicitly compute the transport distance between codebooks by using Wasserstein distance~\cite{hormander2006grundlehren} with Sinkhorn iteration~\cite{cuturi2013sinkhorn}, which formulate as the following equation:
%
\begin{equation}
\begin{aligned}
\mathcal{L}_\text{OT}\left(\mathcal{M}^A, \mathcal{M}^B\right)=\sum\nolimits^d_{i=1}\sum\nolimits^J_{j=1}\left\|\;\mathcal{M}^A_{j,i}-\mathcal{M}^B_{\pi^i({j}),i}\;\right\|_2^2,
\end{aligned}
\label{eq:transport_distance_loss}
\end{equation}
where $J$ denotes the number of samples stored in the memory bank $\mathcal{M}$, $\mathcal{M}_{j,i}$ denote the $i$-th dimension of $j$-th sample in $\mathcal{M}$. The $\pi^i(\cdot)$ is a mapping function to minimize the transport distance of samples between two memory banks as the following:
\begin{equation}
    \pi^i = \mathop{\text{argmin}}_{\pi} \sum\nolimits_{j=1}^{J}\Big\| \mathcal{M}^A_{j, i} - \mathcal{M}^B_{\pi(j), i}\Big\|^2_2.
\end{equation}

Additionally, we minimize the distance between the current quantized feature and the centroid of the corresponding memory bank.
Hence, we use the representative centroid that averages the features of all samples in $\mathcal{M}$ as $\overline{\mathcal{M}}$. The centroids $\overline{\mathcal{M}}$ then is used for measuring the discrepancy with the quantized feature $\tilde{F}$ (defined in Equation~\ref{eq:quantization}), the loss is formulated in the following form:
\begin{equation}
\mathcal{L}_\text{dis}\left(\tilde{F}, \overline{\mathcal{M}}\right)=\left\|\;\tilde{F}-\overline{\mathcal{M}}\;\right\|_2^2.
\label{eq:discrete_transport_distance_loss}
\end{equation}
Similarly, the same operation will be conducted for the abnormal input $Y$.
The overall loss $\mathcal{L}_\text{CDC}$ for optimizing the CDC is:
%
%
%
%
\begin{equation}
\begin{aligned}
    \mathcal{L}_\text{CDC}&=\mathcal{L}_\text{q}(F^A_X, \tilde{F}^A_X) + \mathcal{L}_\text{q}(F^B_Y, \tilde{F}^B_Y) \\ &+ \mathcal{L}_\text{dis}(\tilde{F}^A_X, \overline{\mathcal{M}}^A) + \mathcal{L}_\text{dis}(\tilde{F}^B_Y, \overline{\mathcal{M}}^B)+\mathcal{L}_\text{OT}(\mathcal{M}^A, \mathcal{M}^B),
\end{aligned}
\end{equation}
where $F^\theta_I = \xi^\theta(I)$ and $\tilde{F}^\theta_I = \sigma(F^\theta_I, \mathcal{Z}^\theta, \mathcal{P}^\theta)$ for $\theta \in \{A, B\}$, and $I \in \{X, Y\}$ for paired $(X, Y)$ input.

\subsection{Consistency Deformation Discriminator}
\label{sec:CDD}
The introduction of CDD ensures the reconstructed echocardiogram videos remain consistent in their spatially and temporally visual properties, such as textures and colors. Also, the discriminator acts as an adversary that forces the reconstructed results to conform with the real data in semantic properties, such as structural abnormalities and motion dysfunction of a specific cardiac disease. 
Hence, the CDD consists of two discriminators, denoted as $\eta^{S}(\cdot)$ and $\eta^{T}(\cdot)$, which discriminate reconstructed results and real samples. The $\eta^{S}(\cdot)$ for spatial consistency discriminates every single frame of a video while the $\eta^{T}(\cdot)$ responds to the temporal consistency that takes the whole video as input.
Using the reconstruction process from normal to abnormal as an example, we let $\{\hat{X}_n\}_{n=1}^N = \phi^A(X)$ and $\{\hat{Y}_n\}_{n=1}^N=\hat{Y}$ to represent the reconstructed video and real abnormal video, respectively.
As shown in Fig.~\ref{fig:overview_detail}, globally, we use the $\eta^{T}(\cdot)$ to discriminate the whole reconstructed video $\phi^A(X)$ and sampled real video $\hat{Y}$, as the first term in Equation~\ref{eq:adv_loss}. The $\eta^{T}(\cdot)$ takes each frame of $\phi^A(X)$ and $\hat{Y}$ in order as an image pair for the spatial discrimination as the second term in Equation~\ref{eq:adv_loss}.
%

Locally, we need to guarantee that each region of cardiac can also conduct high-quality reconstruction as well as remain consistent with real cases. For example, for the process of reconstructing normal $X$ to abnormal $\phi^A(X)$, the discrepancy of motion between reconstructed results $\phi^A(X)$ and real abnormal sample $Y$ should remain consistent for a person. 
Hence, we first convert $\hat{Y}$ and $\phi^A(X)$ to non-overlap patches as $\{\hat{Y}_{i,j}\}_{i=1,j=1}^{h,w},\{\hat{X}_{i,j}\}_{i=1,j=1}^{h,w}\in\mathbb{R}^{N\times \frac{H}{h}\times \frac{W}{w}\times 3}$, 
where $\frac{H}{h}, \frac{W}{w} \in \mathbb{Z}^+$, $W$ and $H$ is the width and height of input images, $w$ and $h$ is the size of width and height of the feature map. The overall adversarial loss for global and local discrimination can be formulated as the following Equation: 
%
\begin{equation}
\begin{aligned}
\mathcal{L}_\text{adv}(\phi^A(X),Y)&=\left(\log (\eta^T(\phi^A(X)))+\log (1-\eta^T({Y}))\right)\\
&+\sum\nolimits_{n=1}^{t}\left[\log (1-\eta^S(\hat{X}_n))+\log (\eta^S(\hat{Y}_n))\right]\\
&+\sum\nolimits_{i=1,j=1}^{h,w}\left[\log(1-\eta^T(\hat{X}_{i,j}))+\log(\eta^T(\hat{Y}_{i,j}))\right].
\end{aligned}
\label{eq:adv_loss}
\end{equation}
%
To address the necessity of both global and local discrimination, we conduct the ablation study as shown in Section~\ref{sec:ablation} and Table~\ref{tab:effect_of_module}. For the overall adversarial loss, according to Equation~\ref{eq:adv_loss}, we have $\mathcal{L}_\text{CDD}=\mathcal{L}_\text{adv}(\phi^A(X),Y)+\mathcal{L}_\text{adv}(\phi^B(Y),X)$. In the final, applying the End-to-End training for the CMT and combining the loss from CDC and CDD, the overall loss of our CardiacNet is $\mathcal{L}_{all}=\mathcal{L}_\text{CDC}+\mathcal{L}_\text{CDD}+\mathcal{L}_\text{recon}(X,X^R)+\mathcal{L}_\text{recon}(Y,Y^R)$.

\section{Experiments}
\subsection{Dataset}
We evaluate our method on three datasets, including two public datasets CAMUS~\cite{leclerc2019deep} and Echonet-Dynamic~\cite{ouyang2020video}, as well as our collected dataset  CardiacNet-PAH and CardiacNet-ASD.
\\

\noindent
\textbf{CardiacNet-PAH and CardiacNet-ASD.} We collect datasets from four collaborating hospitals. To guarantee all echocardiogram videos are standards-compliant, each case underwent a video from the apical four-chamber heart (A4C) view are collected, annotated and approved by 5-6 experienced physicians. Ethically, we strictly adhere to the ethical standards of medical research and ensure that the local ethics committee approves all image data collection and experiments. As shown in Table~\ref{tab:detail_dataset}, the CardiacNet-PAH consists of 496 cases for classifying Pulmonary Arterial Hypertension (PAH), and the diagnosis of patients is accessed and approved through Right Heart Catheterization measurement. In CardiacNet-ASD, 231 cases for classifying the Atrial Septal Defect (ASD) are diagnosed and annotated by experienced physicians. 
The resolution of each video is either 800$\times$600 or 1024$\times$768, depending on the type of scanner (Philips or HITACHI). A total of 727 videos are collected, and each video consists of over 100 frames, covering at least two heartbeat cycles.
We also collect Pixel-level annotations of cardiac structure for reconstruction evaluation, including masks for the left ventricle (LV), right ventricle (RV), left atrium (LA), and right atrium (RA) in the A4C view. Five frames are provided with pixel-level annotation masks for each video.
\\

\noindent
\textbf{CAMUS}~\cite{leclerc2019deep} \textbf{and} \textbf{EchoNet-Dynamic}~\cite{ouyang2020video}. CAMUS consists of $500$ echocardiogram videos with pixel-level annotations for the left ventricle, myocardium, and left atrium. EchoNet-Dynamic~\cite{ouyang2020video} (EchoNet) is the largest echocardiogram video dataset, including 10,030 videos. Both datasets annotated 2 frames (end diastole and end systole) of left ventricle segmentation. The Ejection Fraction (EF) score is provided for each video for the regression task. In this paper, we follow the~\cite{mcdonagh20212021} that use cases in CMAUS and EchoNet with EF $\leq50\%$ as the abnormal group while EF $\ge55\%$ as the normal group for classification class.
\subsection{Implementation Details}
\noindent
\textbf{Training.}
The backbone of our methods is built on the generative network~\cite{esser2021taming}. We trained the model using the Adam optimizer with a weight decay of $1e^{-3}$ and a momentum of $0.9$. The model was trained for a total of $1,000$ epochs with an initial learning rate of $2.25e^{-4}$, and the learning rate was decreased by a factor of $0.1$ for every $400$ epochs. The batch size was set to $2$ in our experiment.
For spatial data augmentation, each frame was resized to $144\times 144$ and then randomly cropped to $112\times 112$. The frames were also randomly flipped vertically and horizontally. For temporal data augmentation, we randomly selected $48$ continuous frames from an echocardiogram video and sampled $16$ frames as input equidistantly.
The CardiacNet was split in a ratio of 8:1:1 for training, validation and testing.
For the CAMUS and EchoNet datasets, we follow the same data argumentation recipe as our CardiacNet. We also follow the default dataset split provided by the official setting~\cite{leclerc2019deep} and~\cite{ouyang2020video}. 
\\

\noindent
\textbf{Inference and Testing.}
During this stage, we took the feature extractor $\xi^A(\cdot)$ of network $\phi^A(\cdot)$ saved from the final iteration as our testing model. For cardiac disease assessment tasks, classification and regression, we first freeze the parameter of the feature extractor in the trained model. Then, for each input, we flattened the feature and fine-tuned different tasks with a single Linear layer. 
We report the final results and perform visualization on the testing set. During this stage, we do not conduct any argumentation for input echocardiogram video except resize frames to $144\times 144$ and apply center cropping to $112\times 112$. For the length of input videos in inference, The number of input frames is $16$, and the sample rate is $4$. For evaluating the reconstruction result, we trained a segmentation network according to the segmentation annotation presented by the dataset. During inference, the reconstruction result from pre-trained network $\phi^A(\cdot)$ will be input to the segmentation network and perform the evaluation.
\\

\vspace{-4pt}
\noindent
\textbf{Evaluation Metrics.}
For PAH, ASD and EF classification, we use the Area Under the ROC Curve (AUC) and classification accuracy (ACC) to evaluate the performance of trained networks in classifying anomalies. We predict the EF values and report the Mean Absolute Error (MAE) for CAMUS and Echonet datasets that evaluate the Ejection Fraction (EF) score. In order to evaluate the reconstruction quality, we use Fréchet Inception Distance (FID) to evaluate the quality of recovery images. For ASD, we also introduce the DICE score to evaluate whether recovered images are consistent with the original image in the ventricles and atrium of cardiac structures. This is due to the recovery from ASD to normal does not affect the volume of cardiac structures. For each method, we also compare their efficiency by reporting the inference time, the number of parameters (MParams) and Tera-Flops (TFlops)\footnotetext[2]{One step TFlops of denoising. A total of 1000 steps are used in inference.}. 
\vspace{-8pt}
\subsection{Results}
\vspace{-4pt}
\noindent
\textbf{Result on CardiacNet-PAH and CardiacNet-ASD.} Table~\ref{tab:results_EchoMD} illustrates the comparison result of PAH classification results in CardiacNet-PAH. We currently categorise open-source methods into classification/regression models and reconstruction-based models. The AUC-ROC and ACC illustrate the performance of models in distinguishing normal and abnormal cases. Our CardiacNet achieves $89.32\%$ and $85.71\%$ in AUC-ROC and ACC, respectively, while the HiFuse~\cite{huo2024hifuse} reaches the second-best results with $84.11\%$ and $83.67\%$, where CardiacNet surpass by $+5.21\%$ and $+2.04\%$. Indicating our method can outperform other methods by a considerable margin. For reconstructed image quality evaluation, compared to the Wolleb et al.~\cite{wolleb2022diffusion} reaches $16.12$ in FID score, our method can achieve $14.73$, which shows that our method can perform better reconstruction quality in echocardiogram videos.

Compared to PAH classification, classifying ASD is an easier task due to ASD presenting more significant morphological anomalies. For the classification performance in CardiacNet-ASD, the AUC-ROC and ACC presented by our method are $91.24\%$ and $89.63\%$, outperforming the best baseline DeepGuide~\cite{mallya_2022_deepguide} by a margin of $+6.22\%$ and $+4.84\%$. 
As illustrated in Table~\ref{tab:results_EchoMD}, CardiacNet achieves $15.22$ in FID score with $0.56$ improvement than the second best method Wolleb et al.~\cite{wolleb2022diffusion}. Also, to evaluate that the reconstructed image is consistent in volume sizes of different cardiac structures, our method achieves the best Dice score of $73.52\%$, while other methods are significantly below $70\%$. 
%
%
\begin{table*}[t!]
\setlength{\tabcolsep}{4pt}
\centering
  \caption{The result of classification in our CardiacNet-PAH and CardiacNet-ASD, reporting results in metrics FID, AUC-ROC (\%) and ACC (\%). For ASD, with DICE (\%) score to evaluate the segmentation accuracy of reconstructed images compared with ground truth. The classification networks do not reconstruct the image, and the FID is not provided for these approaches. \underline{Underline} denotes the second-best result.}
  \vspace{-10pt}
  \begin{adjustbox}{width=0.999\linewidth}
    \begin{tabular}{ccccccccccc}
\hlineB{4}
\multicolumn{1}{c|}{\multirow{3}{*}{Methods}} & \multicolumn{7}{c|}{Datasets} & \multicolumn{3}{c}{\multirow{2}{*}{Efficiency}}  \\ \cline{2-8} 
\multicolumn{1}{c|}{} & \multicolumn{3}{c|}{CardiacNet-PAH}     & \multicolumn{4}{c|}{CardiacNet-ASD} \\ \cline{2-11} 
\multicolumn{1}{c|}{} & \multicolumn{1}{c|}{FID$\downarrow$} & \multicolumn{1}{c|}{AUC-ROC$\uparrow$} & \multicolumn{1}{c|}{ACC$\uparrow$}  & \multicolumn{1}{c|}{FID$\downarrow$} & \multicolumn{1}{c|}{DICE$\uparrow$} & \multicolumn{1}{c|}{AUC-ROC$\uparrow$} & \multicolumn{1}{c|}{ACC$\uparrow$} & \multicolumn{1}{c|}{Time$\downarrow$} & \multicolumn{1}{c|}{MParams$\downarrow$} & \multicolumn{1}{c}{TFlops$\downarrow$} \\ \hline \hline
    & \multicolumn{7}{c}{\textit{Classification Network}}     \\ \hline
\multicolumn{1}{c|}{ResNet3D~\cite{hara3dcnns}}  & - & 77.32 & \multicolumn{1}{c|}{71.43} & - & - & 72.25 & \multicolumn{1}{c|}{75.86} & 2.479 & 47.02 & 0.202 \\
\multicolumn{1}{c|}{AGXNet~\cite{yu2022anatomy}} & - & 76.09 & \multicolumn{1}{c|}{72.41} & - & - & 76.52 & \multicolumn{1}{c|}{72.41} & 2.873 & 12.31 & 0.210 \\
\multicolumn{1}{c|}{EchoNet~\cite{ouyang2020video}} & - & 81.63 & \multicolumn{1}{c|}{80.95} & - & - & 83.62 & \multicolumn{1}{c|}{82.75} & 2.653 & 33.19 & 0.848 \\
\multicolumn{1}{c|}{DeepGuide~\cite{mallya_2022_deepguide}} & - & 82.45 & \multicolumn{1}{c|}{81.63} & - & - & \underline{85.02} & \multicolumn{1}{c|}{\underline{84.79}} & 3.780 & 15.60 & 0.748 \\
\multicolumn{1}{c|}{DiffMIC~[40]} & - & 81.73 & \multicolumn{1}{c|}{79.59} & - & - & 82.81 & \multicolumn{1}{c|}{81.48} & 1182 & 88.56 & 38.58$^\dagger$ \\ 
\multicolumn{1}{c|}{HiFuse~\cite{huo2024hifuse}} & - & \underline{84.11} & \multicolumn{1}{c|}{\underline{83.67}} & - & - & 81.08 & \multicolumn{1}{c|}{79.31} & 3.183 & 135.7 & 5.106 \\ \hline 

& \multicolumn{7}{c}{\textit{Reconstruction-Based Methods}}     \\ \hline
\multicolumn{1}{c|}{Vanilla GAN~\cite{esser2021taming}}  & 18.90 & 52.37 & \multicolumn{1}{c|}{46.15} & 19.07 & 63.55 & 60.54 & \multicolumn{1}{c|}{58.62} & 2.221 & 12.95 & 0.842  \\
\multicolumn{1}{c|}{DAE~\cite{kascenas2022denoising}} & 16.39 & 58.91 & \multicolumn{1}{c|}{57.69} & 15.38 & 65.80 & 54.09 & \multicolumn{1}{c|}{53.77} & 1534 & 159.4 & 78.08$^\dagger$ \\
\multicolumn{1}{c|}{VTGAN~\cite{VTGAN2021}} & 17.66 & 58.32 & \multicolumn{1}{c|}{51.72} & 18.10 & 65.13 & 70.92 & \multicolumn{1}{c|}{68.97} & 38.50 & 243.3 & 1.423 \\ 
\multicolumn{1}{c|}{Att. UNet~\cite{schlemper2019attention}} & 18.42 & 57.29 & \multicolumn{1}{c|}{55.17} & 18.95 & 64.30 & 69.81 & \multicolumn{1}{c|}{62.06} & 2.621 & 34.88 & 4.081  \\
\multicolumn{1}{c|}{Wolleb et al.~\cite{wolleb2022diffusion}} & \underline{16.12} & 70.42 & \multicolumn{1}{c|}{67.35} & \underline{15.78} & 68.61 & 67.88 & \multicolumn{1}{c|}{65.51} & 1488 & 89.87 & 45.13$^\dagger$ \\
\multicolumn{1}{c|}{DeScarGAN~\cite{wolleb2020descargan}} & 16.59 & 64.21 & \multicolumn{1}{c|}{71.42} & 17.04 & 68.52 & 71.33 & \multicolumn{1}{c|}{68.97} & 2.756 & 8.528 & 2.756 \\
\multicolumn{1}{c|}{Diff-SCM~\cite{sanchez2022healthy}} & 15.57 & 64.23 & \multicolumn{1}{c|}{61.22} & 16.37 & 63.26 & 69.23 & \multicolumn{1}{c|}{70.83} & 1295 & 53.41 & 40.37$^\dagger$ \\
\multicolumn{1}{c|}{CyTran~\cite{Ristea-CyTran-2023}} & 16.40 & 72.69 & \multicolumn{1}{c|}{69.38} & 16.93 & 70.21 & \underline{74.35} & \multicolumn{1}{c|}{72.41} & 2.769 & 1.191 & 0.125 \\ \hline
\multicolumn{1}{c|}{\textbf{CardiacNet (Ours)}} & \textbf{14.73} & \textbf{89.32} & \multicolumn{1}{c|}{\textbf{85.71}} & \textbf{15.22} & \textbf{73.52} & \textbf{91.24} & \multicolumn{1}{c|}{\textbf{89.63}} & 4.523 & 28.34 & 7.949  \\ \hlineB{4}
\end{tabular}
  \end{adjustbox}
  \label{tab:results_EchoMD}
  \vspace{-10pt}
\end{table*}
\begin{table}[t!]
\setlength{\tabcolsep}{10pt}
\centering
  \caption{The result of ejection fraction regression and abnormal classification in publicly CAMUS~\cite{leclerc2019deep} and EchoNet~\cite{ouyang2020video} dataset. Reporting results in metrics FID, AUC-ROC (\%) and ACC (\%), with Mean Absolute Error (MAE) of ejection fraction score regression for CAMUS and EchoNet. The classification/regression networks do not reconstruct the image, and the FID is not provided for these approaches. \underline{Underline} denotes the second-best result.}
  \vspace{-10pt}
  \begin{adjustbox}{width=0.999\linewidth}
    \begin{tabular}{ccccccccc}
\hlineB{4}
\multicolumn{1}{c|}{\multirow{3}{*}{Methods}} & \multicolumn{8}{c}{Datasets}     \\ \cline{2-9} 
\multicolumn{1}{c|}{} & \multicolumn{4}{c|}{CAMUS}    & \multicolumn{4}{c}{EchoNet}     \\ \cline{2-9} 
\multicolumn{1}{c|}{} & \multicolumn{1}{c|}{FID$\downarrow$} & \multicolumn{1}{c|}{MAE$\downarrow$} & \multicolumn{1}{c|}{AUC$\uparrow$} & \multicolumn{1}{c|}{ACC$\uparrow$}   & \multicolumn{1}{c|}{FID$\downarrow$} & \multicolumn{1}{c|}{MAE$\downarrow$} & \multicolumn{1}{c|}{AUC$\uparrow$} & \multicolumn{1}{c}{ACC$\uparrow$}      \\ \hline \hline
    & \multicolumn{8}{c}{\textit{Classification / Regression Network}}     \\ \hline
\multicolumn{1}{c|}{ResNet3D~\cite{hara3dcnns}} & - & 7.59 & 70.34 & \multicolumn{1}{c|}{68.00} & - & 5.44 & 78.80 & 75.44  \\
\multicolumn{1}{c|}{AGXNet~\cite{yu2022anatomy}} & - & 6.91 & 76.58 & \multicolumn{1}{c|}{72.00} & - & 5.17 & 78.46 & 80.02  \\ 
\multicolumn{1}{c|}{DeepGuide~\cite{mallya_2022_deepguide}} & - & 6.72 & 79.66 & \multicolumn{1}{c|}{74.00} & - & 4.70 & 84.33 & \multicolumn{1}{c}{79.59} \\
\multicolumn{1}{c|}{EchoNet~\cite{ouyang2020video}} & - & \underline{6.30} & \underline{80.75} & \multicolumn{1}{c|}{\underline{76.00}} & - & 4.22 & 83.19 & 81.52  \\
\multicolumn{1}{c|}{HiFuse~\cite{huo2024hifuse}} & - & 6.34 & 80.26 & \multicolumn{1}{c|}{76.00} & - & \underline{4.08} & \underline{85.73} & \multicolumn{1}{c}{\underline{82.41}} \\ \hline
    & \multicolumn{8}{c}{\textit{Reconstruction-Based Methods}}     \\ \hline
\multicolumn{1}{c|}{Vanilla GAN~\cite{esser2021taming}} & 17.24 & 12.59 & 65.11 & \multicolumn{1}{c|}{66.00} & 17.36 & 20.23 & 50.18 & 50.60  \\
\multicolumn{1}{c|}{VTGAN~\cite{VTGAN2021}} & 16.95 & 13.72 & 61.62 & \multicolumn{1}{c|}{56.00} & 15.83 & 12.87 & 61.56 & \multicolumn{1}{c}{61.05} \\ 
\multicolumn{1}{c|}{Att. UNet~\cite{schlemper2019attention}} & 17.72 & 9.48 & 65.60 & \multicolumn{1}{c|}{62.00} & 16.44 & 8.25 & 65.09 & \multicolumn{1}{c}{61.92} \\
\multicolumn{1}{c|}{CyTran~\cite{Ristea-CyTran-2023}} & 15.82 & 8.52 & 66.42 & \multicolumn{1}{c|}{66.00} & 15.07 & 7.59 & 68.45 & \multicolumn{1}{c}{66.53} \\
\multicolumn{1}{c|}{DeScarGAN~\cite{wolleb2020descargan}} & 15.56 & 6.80 & 73.24 & \multicolumn{1}{c|}{68.00} & 14.19 & 7.23 & 73.24 & 71.08  \\
\multicolumn{1}{c|}{Wolleb et al.~\cite{wolleb2022diffusion}} & 15.17 & 8.06 & 75.96 & \multicolumn{1}{c|}{74.00} & \textbf{13.18} & 8.50 & 72.38 & 69.57 \\ \hline
\multicolumn{1}{c|}{\textbf{CardiacNet (Ours)}}& \textbf{14.64} & \textbf{5.97} & \textbf{83.09} & \multicolumn{1}{c|}{\textbf{80.00}} & \underline{13.25} & \textbf{3.83} & \textbf{86.52} & \textbf{84.70}  \\ \hlineB{4}
\end{tabular}
  \end{adjustbox}
  \label{tab:results_camus_echo}
  \vspace{-20pt}
\end{table}
\\

\noindent
\textbf{Result on CAMUS and EchoNet.}
As shown in Table~\ref{tab:results_camus_echo} in columns CAMUS and EchoNet, for the regression task of EF score prediction in both datasets, results achieved by our method are considerably better than others, with $5.97$ and $3.83$ MAE in the regression task. In contrast, the second best method, HiFuse~\cite{huo2024hifuse}, reaches only $6.34$ and $4.08$ in MAE, respectively. Illustrates our method CardiacNet is able to learn the better representation for the regression task. For disease classification, the AUC-ROC and ACC of our method in CMUAS are $83.09\%$ and $79.11\%$, respectively, while reaching $86.52\%$ and $84.70\%$ in EchoNet. The second best method is HiFuse with the AUC-ROC and ACC of $80.26\%$ and $76.13\%$ in CAMUS as well as $85.73\%$ and $82.41\%$ in EchoNet, respectively. Results illustrate our method is more accurate in classifying patients with abnormal left ventricular endocardium in both end-diastole (ED) and end-systole (ES).
Compared with other methods in reconstructing high-quality videos, our method can achieve the FID score of $14.64$ and $13.25$ while Wolleb et al.~\cite{wolleb2022diffusion} achieve $15.17$ and $13.18$ in CMUAS and EchoNet datasets, with the higher reconstruction quality in EchoNet dataset. 
\vspace{-3pt}
\subsection{Ablation Study}
\label{sec:ablation}
\begin{figure}[t!]
    \centering
    \includegraphics[width=0.999\linewidth]{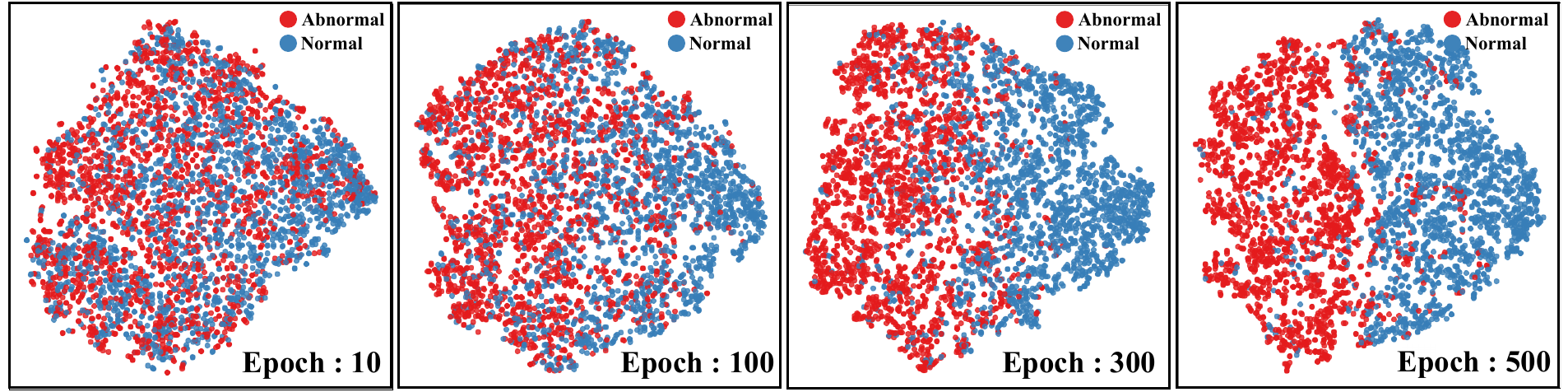}
    \vspace{-18pt}
    \caption{The visualization of t-SNE results between learned embedding of normal and abnormal cases by our CardiacNet in epochs 10, 100, 300 and 500.}
    \label{fig:tsne_visulization}
    \vspace{-4pt}
\end{figure}

\begin{table*}[!t]
    \centering
    \begin{minipage}[t]{.320\linewidth}
      \caption{Effectiveness of CDC and CDD. Results report in CardiacNet-PAH.}
      \vspace{-7pt}
      \setlength{\tabcolsep}{1.9pt}
      \resizebox{0.999\textwidth}{!}{
      \begin{tabular}{c|c|c|c|c}
        \hlineB{2.5}
            \multirow{2}*{CDC} & \multirow{2}*{CDD} & \multicolumn{3}{c}{Results} \\ 
            \cline{3-5}
             & & FID & AUC & ACC \\ \hline \hline
    	\tikzxmark & \tikzxmark & 18.90 & 52.37 & 46.15 \\
            \tikzcmark & \tikzxmark & 16.82 & 80.27 & 79.59 \\
            \tikzxmark & \tikzcmark & 17.09 & 52.46 & 53.84 \\ 
            \tikzcmark & \tikzcmark & \textbf{14.73} & \textbf{89.23} & \textbf{85.71} \\ \hline 
        \hlineB{1.5}
    \end{tabular}
    }
    \label{tab:effect_of_module}
    \end{minipage}
    \hspace{1pt}
    \centering
    \begin{minipage}[t]{.310\linewidth}
      \centering
        \caption{Ablation study of Positional Encoding and Optimal Transport in \textbf{only CDC module}.}
        \vspace{-10pt}
        \setlength{\tabcolsep}{2.7pt}
        \resizebox{0.999\textwidth}{!}{
        \begin{tabular}{c|c|c|c|c}
        \hlineB{3}
	  \multirow{2}*{\shortstack{Pos.\\Encode}} &  \multirow{2}*{\shortstack{Opt.\\Trans}}  &\multicolumn{3}{c}{Results} \\ 
        \cline{3-5} &  & FID & AUC & ACC \\ \hline \hline
        \tikzxmark & \tikzxmark & 18.90 & 52.37 & 46.15 \\
	  \tikzcmark & \tikzxmark & 17.41 & 62.44 & 65.38 \\
        \tikzxmark & \tikzcmark & 18.06 & 78.39 & 75.51 \\
        \tikzcmark & \tikzcmark & \textbf{16.82} & \textbf{80.27} & \textbf{79.59} \\
        \hlineB{3}
	\end{tabular}
    }
    \label{tab:effect_of_CDC}
    \end{minipage}
    \hspace{1pt}
    \begin{minipage}[t]{.320\linewidth}
      \caption{Ablation study of Global and Local discriminator in CDD module (Enabling CDC).}
      \vspace{-10pt}
      \setlength{\tabcolsep}{3.1pt}
      \resizebox{0.999\textwidth}{!}{
      \begin{tabular}{c|c|c|c|c}
        \hlineB{3}
            \multirow{2}*{\shortstack{Global.\\CDD}} & \multirow{2}*{\shortstack{Local.\\CDD}} & \multicolumn{3}{c}{Results} \\ 
            \cline{3-5}
            & & FID & AUC & ACC \\ \hline \hline
            \tikzxmark & \tikzxmark & 16.82 & 80.27 & 79.59 \\
            \tikzcmark & \tikzxmark & 15.62 & 82.41 & 83.67 \\
            \tikzxmark & \tikzcmark & 15.41 & 84.57 & 81.63 \\
            \tikzcmark & \tikzcmark & \textbf{14.73} & \textbf{89.23} & \textbf{85.71} \\
        \hlineB{3}
    \end{tabular}
    }
    \label{tab:effect_of_CDD}
    \end{minipage}%
    \vspace{-18pt}
\end{table*}
\noindent
\textbf{Consistency Deformation Codebook.}
As shown in Table~\ref{tab:effect_of_CDC}, the ablation study of the CDC module consists of positional encoding and optimal transport. Compared to the results of disabling these two modules, the position encoding for temporal consistency can enhance reconstruction quality by around $1.34$ in FID and classification accuracy by around $20\%$. The improvement of the above two numbers contributed by optimal transport is around $0.84$ and $30\%$. These results show both the positional embedding and optimal transport are efficient and can help CardiacNet to learn the better representation of cardiac diseases.
Furthermore, we visualize the embedding features of PAH patients and normal cases in Fig~\ref{fig:tsne_visulization}. Our reconstruction network produces embedding features without using additional layers, which shows that our CDC can help distinguish cardiac structural and motion abnormalities.
\\

\noindent
\textbf{Consistency Deformation Discriminator.}
As shown in Table~\ref{tab:effect_of_CDD}, both global and local discriminators can contribute to the CDD module. Due to their constraint of the spatial and temporal consistency within patches (see section~\ref{sec:CDD}), the CDD brings improvement in reconstructed image quality. Using only the global or local discriminator leads to significant degradation in the FID score and classification accuracy. The ablation study in Table~\ref{tab:effect_of_module} shows the combination of CDD, CDC, and CardiacNet achieves the best performance in both reconstruction and classification. 
\begin{figure}[t!]
    \centering
    \includegraphics[width=0.999\linewidth]{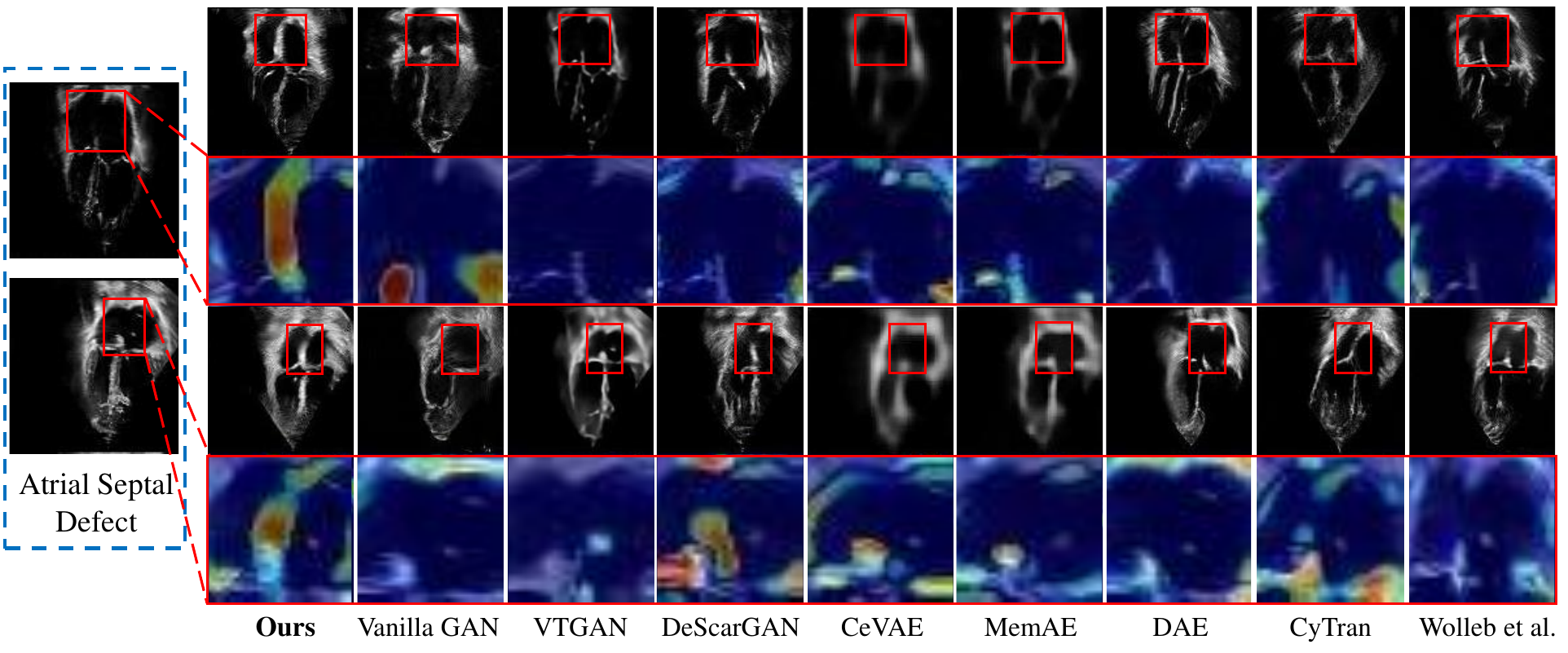}
    \vspace{-18pt}
    \caption{The visualization case of recovery results across ours and eight different reconstruction-based methods~\cite{VTGAN2021,schlegl2019f,wolleb2020descargan,zimmerer2018context,schlemper2019attention,Ristea-CyTran-2023,gong2019memorizing,kascenas2022denoising,wolleb2022diffusion}. We use cases from patients with \textbf{A}trial \textbf{S}eptal \textbf{D}efect (ASD). We let experienced physicians annotate possible abnormal areas and visualize the difference by using the heatmap. (Best view in colour)}
    \vspace{-14pt}
    \label{fig:visulization}
\end{figure}
\\

\noindent
\textbf{The Visualization of Reconstruction Cases.}
As shown in Fig.~\ref{fig:visulization}, our method is able to reconstruct the possible ``normal'' images from abnormal cases compared with other reconstruction methods. Our reconstructed images remain high quality and can provide more reasonable visualization results that are approved by experienced physicians. As shown in two different cases, the disappearance of the atrial septum and the abnormal right atrial volume can be distinguished and recovered while maintaining the reconstruction quality.
\vspace{-8pt}
\section{Conclusion}
\vspace{-4pt}
In this paper, we first proposed a novel CardiacNet for learning the morphological abnormalities and motion dysfunction of cardiac disease through echocardiogram videos. We introduce a new benchmark dataset that includes two different types of cardiac diseases as well as cardiac structure segmentation. All cases are annotated and confirmed by experienced physicians, which can significantly contribute to the medical image analysing community and further the development in detecting morphological abnormalities and motion dysfunction for cardiac diseases. In our future study, we will further our exploration in more fine-grained echocardiogram video reconstruction that enables symptom grading for diseases with the visualization of morphological lesions. Moreover, we will make attempts to involve other state-of-the-art techniques, such as Large language models (LLMs) and multi-modality fusion, to generate more precise and robust results.
%


\newpage
\section*{Acknowledgement}
This work was supported by a research grant from the Beijing Institute of Collaborative Innovation (BICI) under collaboration with the Hong Kong University of Science and Technology under Grant HCIC-004 and a project of Hetao Shenzhen-Hong Kong Science and Technology Innovation Cooperation Zone (HZQB-KCZYB-2020083). 

%
%
\bibliographystyle{splncs04}
\bibliography{main}

\begin{thebibliography}{10}
\providecommand{\url}[1]{\texttt{#1}}
\providecommand{\urlprefix}{URL }
\providecommand{\doi}[1]{https://doi.org/#1}

\bibitem{cuturi2013sinkhorn}
Cuturi, M.: Sinkhorn distances: Lightspeed computation of optimal transport. NIPS  \textbf{26} (2013)

\bibitem{esser2021taming}
Esser, P., Rombach, R., Ommer, B.: Taming transformers for high-resolution image synthesis. In: CVPR. pp. 12873--12883 (2021)

\bibitem{ganame2007regional}
Ganame, J., Mertens, L., Eidem, B.W., Claus, P., D'hooge, J., Havemann, L.M., McMahon, C.J., Elayda, M.A.A., Vaughn, W.K., Towbin, J.A., et~al.: Regional myocardial deformation in children with hypertrophic cardiomyopathy: morphological and clinical correlations. European heart journal  \textbf{28}(23),  2886--2894 (2007)

\bibitem{geske2014deformation}
Geske, J.B., Bos, J.M., Gersh, B.J., Ommen, S.R., Eidem, B.W., Ackerman, M.J.: Deformation patterns in genotyped patients with hypertrophic cardiomyopathy. European Heart Journal--Cardiovascular Imaging  \textbf{15}(4),  456--465 (2014)

\bibitem{ghorbani2020deep}
Ghorbani, A., Ouyang, D., Abid, A., He, B., Chen, J.H., Harrington, R.A., Liang, D.H., Ashley, E.A., Zou, J.Y.: Deep learning interpretation of echocardiograms. NPJ digital medicine  \textbf{3}(1), ~10 (2020)

\bibitem{gong2019memorizing}
Gong, D., Liu, L., Le, V., Saha, B., Mansour, M.R., Venkatesh, S., Hengel, A.v.d.: Memorizing normality to detect anomaly: Memory-augmented deep autoencoder for unsupervised anomaly detection. In: ICCV (2019)

\bibitem{hara3dcnns}
Hara, K., Kataoka, H., Satoh, Y.: Can spatiotemporal 3d cnns retrace the history of 2d cnns and imagenet? In: Proceedings of the IEEE Conference on Computer Vision and Pattern Recognition (CVPR). pp. 6546--6555 (2018)

\bibitem{hormander2006grundlehren}
H{\"o}rmander, F., Totaro, N., Waldschmidt, A.V.M.: Grundlehren der mathematischen wissenschaften 332, vol.~5. Springer (2006)

\bibitem{huo2024hifuse}
Huo, X., Sun, G., Tian, S., Wang, Y., Yu, L., Long, J., Zhang, W., Li, A.: Hifuse: Hierarchical multi-scale feature fusion network for medical image classification. Biomedical Signal Processing and Control  \textbf{87},  105534 (2024)

\bibitem{VTGAN2021}
Kamran, S.A., Hossain, K.F., Tavakkoli, A., Zuckerbrod, S.L., Baker, S.A.: Vtgan: Semi-supervised retinal image synthesis and disease prediction using vision transformers. In: 2021 IEEE/CVF International Conference on Computer Vision Workshops (ICCVW). pp. 3228--3238 (2021). \doi{10.1109/ICCVW54120.2021.00362}

\bibitem{kascenas2022denoising}
Kascenas, A., Pugeault, N., O’Neil, A.Q.: Denoising autoencoders for unsupervised anomaly detection in brain mri. In: International Conference on Medical Imaging with Deep Learning. pp. 653--664. PMLR (2022)

\bibitem{lai2015echocardiography}
Lai, W.W., Mertens, L.L., Cohen, M.S., Geva, T.: Echocardiography in pediatric and congenital heart disease: from fetus to adult. John Wiley \& Sons (2015)

\bibitem{leclerc2019deep}
Leclerc, S., Smistad, E., Pedrosa, J., {\O}stvik, A., Cervenansky, F., Espinosa, F., Espeland, T., Berg, E.A.R., Jodoin, P.M., Grenier, T., et~al.: Deep learning for segmentation using an open large-scale dataset in 2d echocardiography. IEEE transactions on medical imaging  \textbf{38}(9),  2198--2210 (2019)

\bibitem{lin2022echocardiography}
Lin, X., Yang, F., Chen, Y., Chen, X., Wang, W., Chen, X., Wang, Q., Zhang, L., Guo, H., Liu, B., et~al.: Echocardiography-based ai detection of regional wall motion abnormalities and quantification of cardiac function in myocardial infarction. Frontiers in Cardiovascular Medicine  \textbf{9},  903660 (2022)

\bibitem{lin2023learning}
Lin, Y., Luo, Z., Zhao, W., Li, X.: Learning deep intensity field for extremely sparse-view cbct reconstruction. In: Medical Image Computing and Computer Assisted Intervention -- MICCAI 2023. pp. 13--23. Springer Nature Switzerland (2023)

\bibitem{lin2024learning3dgaussiansextremely}
Lin, Y., Wang, H., Chen, J., Li, X.: Learning 3d gaussians for extremely sparse-view cone-beam ct reconstruction (2024), \url{https://arxiv.org/abs/2407.01090}

\bibitem{lin2024c2rv}
Lin, Y., Yang, J., Wang, H., Ding, X., Zhao, W., Li, X.: C{\textasciicircum}2rv: Cross-regional and cross-view learning for sparse-view cbct reconstruction. In: Proceedings of the IEEE/CVF Conference on Computer Vision and Pattern Recognition (CVPR). pp. 11205--11214 (June 2024)

\bibitem{liu2023deep}
Liu, B., Chang, H., Yang, D., Yang, F., Wang, Q., Deng, Y., Li, L., Lv, W., Zhang, B., Yu, L., et~al.: A deep learning framework assisted echocardiography with diagnosis, lesion localization, phenogrouping heterogeneous disease, and anomaly detection. Scientific Reports  \textbf{13}(1), ~3 (2023)

\bibitem{lu2022yolox}
Lu, Y., Li, K., Pu, B., Tan, Y., Zhu, N.: A yolox-based deep instance segmentation neural network for cardiac anatomical structures in fetal ultrasound images. IEEE/ACM Transactions on Computational Biology and Bioinformatics  (2022)

\bibitem{mallya_2022_deepguide}
Mallya, M., Hamarneh, G.: Deep multimodal guidance for medical image classification. In: MICCAI. Springer (2022)

\bibitem{mcdonagh20212021}
McDonagh, T.A., Metra, M., Adamo, M., Gardner, R.S., Baumbach, A., B{\"o}hm, M., Burri, H., Butler, J., {\v{C}}elutkien{\.e}, J., Chioncel, O., et~al.: 2021 esc guidelines for the diagnosis and treatment of acute and chronic heart failure: Developed by the task force for the diagnosis and treatment of acute and chronic heart failure of the european society of cardiology (esc) with the special contribution of the heart failure association (hfa) of the esc. European heart journal  \textbf{42}(36),  3599--3726 (2021)

\bibitem{mcleod2018echocardiography}
Mcleod, G., Shum, K., Gupta, T., Chakravorty, S., Kachur, S., Bienvenu, L., White, M., Shah, S.B.: Echocardiography in congenital heart disease. Progress in cardiovascular diseases  \textbf{61}(5-6),  468--475 (2018)

\bibitem{meena2023weakly}
Meena, T., Kabiraj, A., Reddy, P.B., Roy, S.: Weakly supervised confidence aware probabilistic cam multi-thorax anomaly localization network. In: 2023 IEEE 24th International Conference on Information Reuse and Integration for Data Science (IRI). pp. 309--314. IEEE (2023)

\bibitem{niemann2012echocardiographic}
Niemann, M., Liu, D., Hu, K., Cikes, M., Beer, M., Herrmann, S., Gaudron, P.D., Hillenbrand, H., Voelker, W., Ertl, G., et~al.: Echocardiographic quantification of regional deformation helps to distinguish isolated left ventricular non-compaction from dilated cardiomyopathy. European journal of heart failure  \textbf{14}(2),  155--161 (2012)

\bibitem{ouyang2020video}
Ouyang, D., He, B., Ghorbani, A., Yuan, N., Ebinger, J., Langlotz, C.P., Heidenreich, P.A., Harrington, R.A., Liang, D.H., Ashley, E.A., et~al.: Video-based ai for beat-to-beat assessment of cardiac function. Nature  \textbf{580}(7802),  252--256 (2020)

\bibitem{oxborough2012right}
Oxborough, D., Sharma, S., Shave, R., Whyte, G., Birch, K., Artis, N., Batterham, A.M., George, K.: The right ventricle of the endurance athlete: the relationship between morphology and deformation. Journal of the American Society of Echocardiography  \textbf{25}(3),  263--271 (2012)

\bibitem{popp1976echocardiographic}
Popp, R.L.: Echocardiographic assessment of cardiac disease. Circulation  \textbf{54}(4),  538--552 (1976)

\bibitem{pu2024hfsccd}
Pu, B., Li, K., Chen, J., Lu, Y., Zeng, Q., Yang, J., Li, S.: Hfsccd: a hybrid neural network for fetal standard cardiac cycle detection in ultrasound videos. IEEE Journal of Biomedical and Health Informatics  (2024)

\bibitem{pu2022mobileunet}
Pu, B., Lu, Y., Chen, J., Li, S., Zhu, N., Wei, W., Li, K.: Mobileunet-fpn: A semantic segmentation model for fetal ultrasound four-chamber segmentation in edge computing environments. IEEE Journal of Biomedical and Health Informatics  \textbf{26}(11),  5540--5550 (2022)

\bibitem{puunsupervised}
Pu, B., Lv, X., Yang, J., Guannan, H., Dong, X., Lin, Y., Shengli, L., Ying, T., Fei, L., Chen, M., et~al.: Unsupervised domain adaptation for anatomical structure detection in ultrasound images. In: Forty-first International Conference on Machine Learning

\bibitem{pu2024m3}
Pu, B., Wang, L., Yang, J., He, G., Dong, X., Li, S., Tan, Y., Chen, M., Jin, Z., Li, K., et~al.: M3-uda: A new benchmark for unsupervised domain adaptive fetal cardiac structure detection. In: Proceedings of the IEEE/CVF Conference on Computer Vision and Pattern Recognition. pp. 11621--11630 (2024)

\bibitem{pu2021fetal}
Pu, B., Zhu, N., Li, K., Li, S.: Fetal cardiac cycle detection in multi-resource echocardiograms using hybrid classification framework. Future Generation Computer Systems  \textbf{115},  825--836 (2021)

\bibitem{Ristea-CyTran-2023}
Ristea, N.C., Miron, A.I., Savencu, O., Georgescu, M.I., Verga, N., Khan, F.S., Ionescu, R.T.: Cytran: Cycle-consistent transformers for non-contrast to contrast ct translation. Neurocomputing  (2023). \doi{10.1016/j.neucom.2023.03.072}

\bibitem{ryser2022anomaly}
Ryser, A., Manduchi, L., Laumer, F., Michel, H., Wellmann, S., Vogt, J.E.: Anomaly detection in echocardiograms with dynamic variational trajectory models. In: Machine Learning for Healthcare Conference. pp. 425--458. PMLR (2022)

\bibitem{sanchez2022healthy}
Sanchez, P., Kascenas, A., Liu, X., O’Neil, A.Q., Tsaftaris, S.A.: What is healthy? generative counterfactual diffusion for lesion localization. In: MICCAI Workshop on Deep Generative Models. pp. 34--44. Springer (2022)

\bibitem{sanjeevi2023automatic}
Sanjeevi, G., Gopalakrishnan, U., Pathinarupothi, R.K., Madathil, T.: Automatic diagnostic tool for detection of regional wall motion abnormality from echocardiogram. Journal of medical systems  \textbf{47}(1), ~13 (2023)

\bibitem{schafer2023myocardial}
Sch{\"a}fer, M., Mitchell, M.B., Frank, B.S., Barker, A.J., Stone, M.L., Jaggers, J., von Alvensleben, J.C., Hunter, K.S., Friesen, R.M., Ivy, D.D., et~al.: Myocardial strain-curve deformation patterns after fontan operation. Scientific Reports  \textbf{13}(1),  11912 (2023)

\bibitem{schlegl2019f}
Schlegl, T., Seeb{\"o}ck, P., Waldstein, S.M., Langs, G., Schmidt-Erfurth, U.: f-anogan: Fast unsupervised anomaly detection with generative adversarial networks. Medical image analysis  \textbf{54},  30--44 (2019)

\bibitem{schlemper2019attention}
Schlemper, J., Oktay, O., Schaap, M., Heinrich, M., Kainz, B., Glocker, B., Rueckert, D.: Attention gated networks: Learning to leverage salient regions in medical images. Medical image analysis  \textbf{53},  197--207 (2019)

\bibitem{silva2022constrained}
Silva-Rodr{\'\i}guez, J., Naranjo, V., Dolz, J.: Constrained unsupervised anomaly segmentation. Medical Image Analysis  \textbf{80},  102526 (2022)

\bibitem{sun2023chamber}
Sun, D., Hu, Y., Li, Y., Yu, X., Chen, X., Shen, P., Tang, X., Wang, Y., Lai, C., Kang, B., et~al.: Chamber attention network (can): Towards interpretable diagnosis of pulmonary artery hypertension using echocardiography. Journal of Advanced Research  (2023)

\bibitem{tseng2024real}
Tseng, C.H., Chien, S.J., Wang, P.S., Lee, S.J., Pu, B., Zeng, X.J.: Real-time automatic m-mode echocardiography measurement with panel attention. IEEE Journal of Biomedical and Health Informatics  (2024)

\bibitem{upton1976echocardiographic}
Upton, M., Gibson, D., Brown, D.: Echocardiographic assessment of abnormal left ventricular relaxation in man. Heart  \textbf{38}(10),  1001--1009 (1976)

\bibitem{van2017neural}
Van Den~Oord, A., Vinyals, O., et~al.: Neural discrete representation learning. NIPS  \textbf{30} (2017)

\bibitem{wolleb2022diffusion}
Wolleb, J., Bieder, F., Sandk{\"u}hler, R., Cattin, P.C.: Diffusion models for medical anomaly detection. In: MICCAI. pp. 35--45. Springer (2022)

\bibitem{wolleb2020descargan}
Wolleb, J., Sandk{\"u}hler, R., Cattin, P.C.: Descargan: Disease-specific anomaly detection with weak supervision. In: MICCAI. pp. 14--24. Springer (2020)

\bibitem{yang2023graphecho}
Yang, J., Ding, X., Zheng, Z., Xu, X., Li, X.: Graphecho: Graph-driven unsupervised domain adaptation for echocardiogram video segmentation. In: Proceedings of the IEEE/CVF International Conference on Computer Vision. pp. 11878--11887 (2023)

\bibitem{yu2022anatomy}
Yu, K., Ghosh, S., Liu, Z., Deible, C., Batmanghelich, K.: Anatomy-guided weakly-supervised abnormality localization in chest x-rays. In: MICCAI. pp. 658--668. Springer (2022)

\bibitem{zaman2021spatio}
Zaman, F., Ponnapureddy, R., Wang, Y.G., Chang, A., Cadaret, L.M., Abdelhamid, A., Roy, S.D., Makan, M., Zhou, R., Jayanna, M.B., et~al.: Spatio-temporal hybrid neural networks reduce erroneous human “judgement calls” in the diagnosis of takotsubo syndrome. EClinicalMedicine  \textbf{40} (2021)

\bibitem{zhang2020viral}
Zhang, J., Xie, Y., Pang, G., Liao, Z., Verjans, J., Li, W., Sun, Z., He, J., Li, Y., Shen, C., et~al.: Viral pneumonia screening on chest x-rays using confidence-aware anomaly detection. IEEE transactions on medical imaging  \textbf{40}(3),  879--890 (2020)

\bibitem{zheng2023gl}
Zheng, Z., Yang, J., Ding, X., Xu, X., Li, X.: Gl-fusion: Global-local fusion network for multi-view echocardiogram video segmentation. In: International Conference on Medical Image Computing and Computer-Assisted Intervention. pp. 78--88. Springer (2023)

\bibitem{zimmerer2018context}
Zimmerer, D., Kohl, S.A., Petersen, J., Isensee, F., Maier-Hein, K.H.: Context-encoding variational autoencoder for unsupervised anomaly detection. arXiv preprint arXiv:1812.05941  (2018)

\end{thebibliography}
\end{document}